%% file: rpki_mpc.tex
\documentclass[10pt, conference, letterpaper]{IEEEtran}

\usepackage{amsmath}
\usepackage{amssymb}
\usepackage{color}
\usepackage{todonotes}
\usepackage{balance}
\usepackage{url}
\usepackage{stmaryrd}
\usepackage{booktabs}
\usepackage{framed}
\usepackage{xparse}

\usepackage{slashbox}

\usepackage{datetime}
\usepackage{tikz}
\usepackage{pgfplots}
\usetikzlibrary{backgrounds, calc, fit, positioning, shapes.geometric}

\newcommand\kgen{\ensuremath{\mathsf{KGen}}}
\newcommand\secparam{\ensuremath{1^{\lambda}}}
\newcommand\sk{\ensuremath{\mathsf{sk}}}
\newcommand\pk{\ensuremath{\mathsf{pk}}}
\newcommand\sig{\ensuremath{\mathsf{Sig}}}
\newcommand\Zp{\ensuremath{\mathbb{Z}_{p}}}
\newcommand\verify{\ensuremath{\mathsf{Vf}}}

\newcommand\pick{\ensuremath{\leftarrow}}
\newcommand{\from}{\pick}
\newcommand\sample{\ensuremath{\from}}

\newcommand\sshare[1]{[{#1}]}
\newcommand\gshare[1]{\langle{#1}\rangle}

\newcommand\Open{\mathsf{Open}}

\NewDocumentEnvironment{Boxed}{mmmo}
{%
	\begin{figure}
		\begin{framed}
			\vspace{-0.5em}\hspace{-0.5em}\textbf{#1}
		}{%
		\end{framed}
		\caption{#3}
		\label{#2}
	\end{figure}
}

\title{Privacy Preserving and Resilient RPKI}

\author{\IEEEauthorblockN{Kris Shrishak}
	\IEEEauthorblockA{\textit{Technical University Darmstadt}\\
		Darmstadt, Germany \\
	}
	\and
	\IEEEauthorblockN{Haya Shulman}
	\IEEEauthorblockA{\textit{Fraunhofer SIT} \\
		Darmstadt, Germany \\
	}
}
\begin{document}
	
	\maketitle
	
	\thispagestyle{plain}
	\pagestyle{plain}
	
	\input{abstract}	
	\input{introduction}

	\input{preliminaries}

	%\input{system}
	\input{distributed_rpki}

	\input{eval}

	\input{analysis}

\input{relatedworks}

	\input{conclusion}

\input{acknowledgement}
	%\newpage
	
%-------------------------------------------------------------------------------
\bibliographystyle{plain}
\balance
%\bibliography{reference}
\bibliography{ref}

%%%%%%%%%%%%%%%%%%%%%%%%%%%%%%%%%%%%%%%%%%%%%%%%%%%%%%%%%%%%%%%%%%%%%%%%%%%%%%%%
\end{document}

%% file: abstract.tex
\begin{abstract}
	
Resource Public Key Infrastructure (RPKI) is vital to the security of inter-domain routing. However, RPKI enables Regional Internet Registries (RIRs) to unilaterally takedown IP prefixes - indeed, such attacks have been launched by nation-state adversaries. The threat of IP prefix takedowns is one of the factors hindering RPKI adoption.  

%The deployment rate of RPKI has been low and we argue in this work that addressing the power asymmetry can assist in a more trustworthy system.

In this work, we propose the first distributed RPKI system, based on threshold signatures, that requires the coordination of a number of RIRs to make changes to RPKI objects; hence, preventing unilateral prefix takedown. We perform extensive evaluations using our implementation demonstrating the practicality of our solution. Furthermore, we show that our system is scalable and remains efficient even when RPKI is widely deployed.

\end{abstract}

%% file: introduction.tex
\section{Introduction}

Resource Public Key Infrastructure (RPKI)~\cite{DBLP:journals/rfc/rfc6480} is a cryptographic method to secure inter-domain routing against prefix and sub-prefix hijacks.
It is also a prerequisite for Border Gateway Protocol Security (BGPsec)~\cite{DBLP:journals/rfc/rfc8205}.
In RPKI, Regional Internet Registries (RIRs) allocate IP prefixes and authorize specific autonomous systems (ASes) to be the origin of routes.
This information is stored in route origin authorization (ROA).
Routers use the ROAs to distinguish legitimate routes from leaked or hijacked routes.
This is known as route origin validation (ROV).

The insecurity of inter-domain routing and the ability of RPKI to address the insecurity has not transpired into wide-scale deployment of RPKI~\cite{gilad2017we,DBLP:conf/dsn/HlavacekHSW18}.
One of the reasons is the possibility of RIRs to unilaterally takedown IP prefixes, either deliberately or accidentally, that will result in the prefix of the affected ASes being unreachable when ROV is performed~\cite{DBLP:conf/hotnets/CooperHBRG13}.
The hierarchical structure of RPKI gives RIRs the power to revoke and invalidate any objects that it has issued.

As centralized authorities are easy targets for legal surveillance and coercion, is it possible to prevent a state-sponsored attacker from imposing its demands on RIRs without drastically changing the structure of RPKI?
The RIRs are bound by the law of the country they are based in.
Their members who are based in different countries do not have a recourse when their prefix is taken down.

In the past, there have been situations where these problems have taken practical relevance.
In 2011, RIPE NCC took the state of Netherlands to court when the Dutch police ordered to it to lock registration of four IP address blocks~\cite{ripe2013,ripe2011a}.
Nevertheless, it was forced to lock down the registrations.
%In 2014, a court in the United States of America issued a writ of attachment on the country-level top-level domain of Iran, Syria and North Korea and the associated IP addresses~\cite{icann2014}.
%Such demands from the court are possible due to the centralization of power and the hierarchical structure of systems.
More recently, RIPE NCC mistakenly deleted 2669 ROAs on 1 April 2020 and were reinstated on 2 April~\cite{ripe2020a}. This meant that the announcement for these resources were `unknown'. On the day when these ROAs were missing, RosTelecom had a route leak~\cite{Qrator}. While the two events seem independent, according to RIPE NCC, 12 prefixes whose ROAs were deleted were affected by the route leak.
Furthermore, RIPE NCC transferred an IP prefix block from a member to another entity based on a German court order transferred to them through a Dutch court%. RIPE NCC states that ``the rights to the registration could be seized for the recovery of money''
~\cite{ripe2020b}. As a matter of procedure, they will do the same if similar situations arises in the future. In the context of RPKI, this means RIPE NCC will ``revoke any certificates generated by the RIPE NCC''~\cite{ripe2019}.

%One approach that has been proposed to address the issue of disproportionate power in the hands of RPKI authorities is to add transparency logs and dead objects to RPKI to note the consent of the INR owner for revocation~\cite{DBLP:conf/sigcomm/HeilmanCRG14,LZLS19}.
%Another approach is to replace the existing RPKI system by using blockchains~\cite{DBLP:conf/hotnets/HariL16}.
%This approach eliminates the possibility of RPKI authorities revoking previously allocated resource while they remain part of the blockchain by providing new resources.
%%However, both these approaches do not account for the cases where there might be a legitimate reason to revoke the allocated IP-address blocks, e.g., due to fraud by INR owner~\cite{Arin19}.
%%It should be possible to revoke allocations under legitimate circumstances, but not otherwise.
%Th former approach requires effort not only from the CAs but also from relying parties.
%The use of blockchain in the latter approach raises other deployment issues such as consensus algorithm and incentive for the nodes to run the blockchain.
%If Proof-of-Stake is used as the consensus algorithm, as proposed in \cite{DBLP:conf/ithings/PaillisseFGLPLK18}, then the nodes with greater stake, e.g., large providers who are allocated large subsets of IP addresses will become powerful players, which will create another form of power imbalance.

In this work we address these issues that are prevalent in the deployed RPKI system by constructing a distributed RPKI system that relies on threshold signatures, a specific instance of secure multiparty computation (MPC).
Our solution, without requiring significant changes to BGP and RPKI, restricts the power of RIRs and only allows revocation of allocated resources in legitimate cases with the cooperation of a number of RIRs.
%We limit access of key material to RPKI authorities while leaving open the possibility to revoke previous allocated resources in legitimate cases.
%Rather than trusting each RIR completely, we propose to use MPC to sign and revoke resource objects using a distributed protocol.

\subsection{Significance of the threat model}

BGP without RPKI operates in a default-accept mode where any autonomous system (AS) can announce a BGP route for any IP prefix and the other ASes will accept the route by default.
The default-accept mode has made BGP vulnerable to prefix hijacks, where a malicious AS announces a route for IP prefixes it does not own such that the traffic for those prefixes are sent to it, and sub-prefix hijacks, where a malicious AS announces a more specific IP prefix than the one that has been allocated ~\cite{DBLP:conf/sigcomm/BallaniFZ07,DBLP:journals/pieee/ButlerFMR10,Cowie10,DBLP:journals/comsur/HustonRA11}.

% Amazon (AS16509) Route53 hijack  –Apr 2018–AS10279 (eNET) announced/originated more specifics (/24s) of Amazon Route53’s prefix (205.251.192.0/21)
% -- 205.251.192.0/24 ....... 205.251.199.0/24 (https://ip-ranges.amazonaws.com/ip-ranges.json)
% During the period, DNS servers in the hijacked range only responded to queries for myetherwallet.com
% Responded with addresses associated with AS41995/AS48693

% https://www.manrs.org/2019/06/large-european-routing-leak-sends-traffic-through-china-telecom/

% Internal numbering leak out: https://twitter.com/DougMadory/status/1225486918609227777

%RPKI has been designed to secure BGP against such attacks.
%RPKI authorities attest to the allocation of IP-addresses to AS numbers while routers use the attestation to distinguish valid routes from invalid routes.
%Not only does RPKI avoid making changes to BGP, it also does not require online cryptography during routing.
%%Moreover, BGPsec, an extension of BGP, that provides path-validation relies on RPKI for its functioning~\cite{DBLP:journals/rfc/rfc8205}.
%Considering the benefits of RPKI, it is important to question whether it has a strong enough threat model.

RPKI entrusts hierarchical and centralized authorities to be honest.
Malfunctions or coercion by law enforcement authorities is not incorporated into the threat model.
Such a weak threat model creates an imbalance of power between the RIRs and its members.
Moreover, the power imbalance with RPKI is greater than with Web PKI.
In RPKI, there is no option to request certificates from different authorities.
Hence, the reliance on specific RIRs is greater.

Members are further weakened when the authority is based in a different country than their own.
The manipulations at the level of BGP is more coarse-grained than domain name seizures as BGP granularity is limited to $ /24 $, i.e., 256 IPv4 addresses~\cite{DBLP:conf/hotnets/CooperHBRG13}.
The RIRs are bound by the law of the country they are based in.
If members are affected, they may need to take the issue up in another country.
The slow process may result in a loss of business.

\subsection{Threshold signatures for RPKI}

We propose a distributed RPKI system based on threshold signatures.
Threshold signatures is a cryptographic technique where a threshold of $ t + 1 $ parties out of a set of $ n $ parties are required to jointly compute a signature on a message.
Although the signing process is distributed, the verification scheme remains unchanged.
Threshold signatures are more robust to adversarial attacks in settings where signatures are to be generated in a system where individual parties cannot be entirely trusted.

Threshold signatures provide a method to distribute trust and they are practical in settings where the number of participating parties is small.
One of the deployment method of RPKI is \emph{hosted} RPKI.
There are five RIRs and threshold signature protocols are practical when there are only five participating parties.
Hence, our system requires a threshold of them to agree before making changes to RPKI objects.
This prevents any RIR from unilaterally making changes.

Our solution can be described as follows: threshold signatures use shares of the private key, where each of the five RIRs will have a share of the private key while none of them have the entire private key.
Using only the shares, the RIRs can collaboratively sign ROAs and CRLs.
However, they cannot unilaterally perform any of these actions.
Our mechanism prevents them from acting maliciously unilaterally.
Most importantly, threshold signatures support a stronger threat model where corrupted RPKI authorities are not entirely trusted and yet play a significant role in making BGP secure.

%\kris[inline]{Note the economic feasibility of your design. What are the incentives for RIRs to participate in this?}

{\bf Contributions.} A summary of our contributions:

$\bullet$ We construct a distributed RPKI system based on threshold signatures that addresses three issues: (1) preventing unilateral IP prefix takedowns, (2) limiting the scope and implications of attacks on RIRs, and (3) enabling validation in case of missing trust anchor.
	
$\bullet$ We propose two deployment models of our solution and discuss the trade-offs in these models.
	
$\bullet$ We show the performance of our distributed RPKI system based on four threshold signature protocols, all of which have a stronger threat model than the existing RPKI system.

$\bullet$ We perform extensive evaluation of our system and show that our system is not only efficient for today's requirements, it can also meet future demands.

{\bf Outline.} We provide preliminaries in Section~\ref{sec:preliminaries}.
We elaborate on the system and threat model and describe our distributed RPKI system % and   
in Section~\ref{sec:system}.
Then we discuss the performance of our distributed RPKI system in Section~\ref{sec:eval}.
In Section~\ref{sec:analysis}, we analyse historical RPKI data to understand the number of ROAs issued/revoked over time and show that our system satisfies the requirements.
Finally, we discuss the related works in Section~\ref{sec:relatedworks} and we conclude in Section~\ref{sec:conclusion}.

%\kris[inline]{This work does not raise any ethical issues}

% RIRs can terminate resources allocated to members under certain conditions.
% However, the members can only state their claim in a specific country which may not be their own.
% In the case of RIPE, it is Dutch court.
% Similarly, RIPE only accepts law enforcement requests through Dutch authorities and not others.
% It is easier for other states to send a warrant through Dutch authorities than it is for RIPE members.

%It must not be possible to unilaterally revoke allocated Ip-prefix.
%It should be easy to deploy without requiring wide-spread changes to the existing infrastructure.

%% file: preliminaries.tex
\section{Preliminaries}

\label{sec:preliminaries}\label{sec:ecdsa}

%\kris[inline]{Emphasize offline crypto}
%
%\kris[inline]{Mention the hierarchical structure of RPKI before discussing the benefits of RPKI}
% https://cs-people.bu.edu/heilman/sigRPKI.pdf
%
%Resource Public Key Infrastructure (RPKI)~\cite{DBLP:journals/rfc/rfc6480} is a cryptographic method where resource owners such as Autonomous systems (ASes) issue signatures which allow routers to check whether the routes that they received is from an AS (using Route Origin Authorizations (ROAs)) who is allowed to update routes for this IP prefix.
%By checking the validity of the route update, honest misconfiguration as well as prefix hijacking attacks can be prevented.
%
%%\kris[inline]{Maybe write a short paragraph on CA certificate and End-entity certificate as they are the ones that are revoked and not the ROAs themselves}
%{\bf RPKI.} 

\subsection{RPKI}
RPKI architecture includes CA certificates, end-entity (EE) certificates and trust anchor.
A resource holder needs a CA certificate to sub-allocate resources and to issue resource certificates. 
EE certificates verify signed objects (e.g., ROAs and manifests).
The private key corresponding to the public key in an EE certificate cannot be used to sign other certificates.
There is a one-to-one mapping between EE certificate and signed objects. If the EE certificate is revoked, then the corresponding signed object is automatically revoked.
CA certificate is used to sign EE certificate.
A trust anchor is a self-signed X.509 CA certificate in RPKI that is at the head of the chain and it is assumed to be trusted.
In X. 509 architecture, the chain of trust is derived from this authoritative certificate.
The trust anchor contains a public key in the \texttt{subjectPublicKeyInfo} field along with the associated data that are used by the relying parties to validate a signature on a certificate or signed objects, such as ROAs~\cite{DBLP:journals/rfc/rfc8630, DBLP:journals/rfc/rfc6024}

%The entity holding this certificate has to be trusted for the validation of the other certificates in the hierarchy.
%\kris[inline]{Maybe include a figure}
%
%An AS is a set of Internet routable IP prefixes belonging to a network or a collection of networks that are all managed by a single entity.
%An AS utilises a common routing policy controlled by the entity and is identified by a globally unique 16 or 32-bit number.
%The AS number (ASN) as well as IP address blocks are assigned by one of the five Regional Internet Registries (RIRs).
%

ROAs are digitally signed objects, X.509 certificates~\cite{DBLP:journals/rfc/rfc3779,DBLP:journals/rfc/rfc5280}, that provide a method to verify that an IP address block holder (RIR) has authorized an AS to originate routes to specific prefixes within that address block.
Note that each ROA includes exactly one ASN. However, multiple ASNs may be authorized, but each one requires a separate ROA.
Moreover, issuance of subordinate certificates corresponds to sub-allocation of IP-addresses.
A Certificate Revocation List (CRL) is a list of resource certificates that have been revoked, and should not be relied upon by the relying parties.
A CRL is always issued by the same CA that issues the corresponding certificates.

There are two RPKI models: delegated RPKI and hosted RPKI.
In the delegated RPKI model, AS runs a CA as a child of RIR (or NIR or LIR), generates its own certificate, gets it signed by the parent CA.
This model allows the AS to operate independent of the parent RIR.
For large operators of a global network, this model is suitable so that they do not need to maintain ROAs through the different web interfaces of the RIRs.
However, this model is not suitable for all as it requires running a CA and maintaining the ROAs.

In the hosted-RPKI model, RIRs host the CA, that is, the same entity that allocates IP resources also runs the CA to validate the ROAs.
Thus, in this model, they are trust anchors.
In a way, this is meaningful as the RIRs already know the owner of the address space.
Existing RPKI systems are tied-up with the login credentials of the ASes at the RIR.
Signing and key rollover is automatic.
It is easy for the owners of the address space to begin using hosted RPKI than delegated RPKI as the CA functionality is taken care of by the RIR.
This model is convenient for most ASes.
%, and this is the only option provided by LACNIC~\footnote{As of late 2019.}.
It is easier to use and it is especially useful for members with a small network and with limited resources.
Even large providers such as Cloudflare make use of hosted RPKI~\footnote{\url{https://ripe77.ripe.net/presentations/156-RPKI-deployment-at-scale-RIPE-1.pdf}}.
Furthermore, the RIR assumes responsibility to publish the signed objects.
%Note that the RIR holds the private key in this case.
However, this convenience comes at the cost of further centralization of power as the RIRs also handle the private keys used to sign ROAs.

\subsection{ECDSA}
%\label{sec:ecdsa}

%{\bf ECDSA.} 
The scheme is parameterized by a curve point $G$ of prime order $p$, and we write $\Zp$ for the field of order $p$.
We use $H$ to denote a function mapping arbitrary length messages unto elements of $\Zp$.

\noindent \textbf{Key Generation}.	$\kgen(\secparam)$
	\begin{enumerate}
		\item Sample at random $\sk\from\Zp$ as the signing key.
		\item Compute $\pk=\sk\cdot G$ as the public verification key.
		\item Output $(\sk,\pk)$.
	\end{enumerate}
	
\noindent	\textbf{Signing}. $\sig(\sk, M)$
	\begin{enumerate}
		\item Sample at random an instance key $k\from\Zp$.
		\item Compute $R = (r_{x},r_{y})=k\cdot G$. If $r_{x}\equiv 0\pmod{p}$, go back to step 1.
		\item Compute $s=k^{-1}(H(M) + \sk\cdot r_{x})$ where $H$ is a hash function.
		\item Output $\sigma=(r_{x}, s)$.
	\end{enumerate}

\noindent	\textbf{Verification}. $\verify(\pk, M, \sigma)$
	\begin{enumerate}
		\item Let $(r_{x}',r_{y}')=s^{-1}(H(M)\cdot G + r_{x}\cdot \pk)$.
		\item Output $r_{x}'=r_{x}$.
	\end{enumerate}

\noindent \textbf{Correctness}.
We have
\begin{align}
s^{-1} &(H(M)\cdot G +r_{x} \cdot \pk)  \\
&= k(H(M) + \sk\cdot r_{x})^{-1}(H(M)\cdot G + r_{x}\cdot \pk) \\
&= k\cdot G\cdot ((H(M) + \sk\cdot r_{x})^{-1}(H(M) + \sk\cdot r_{x})) \\
&= k\cdot G = (r_{x},r_{y}),
\end{align}
which shows that valid signatures verify.

%\kris[inline]{Explain why we use ECDSA}

\subsection{MPC and threshold signatures}

%{\bf MPC and threshold signatures.} 
In our work, we use threshold signature protocols that are based on secret sharing.
Specifically, we use additive sharing and Shamir sharing schemes.
We use the notation $ \sshare{a} $ to denote a value $ a $ that is secret-shared, that is, no single party can access it.
For $a\in\Zp$, the shares $ \sshare{a} $ are also elements of $ \Zp $.
%We use $ \gshare{a} $ to denote the secret shares of $ a \cdot G $.
%We use the command $ \Open $ to reconstruct from the secret shared values such that $a\from \Open(\sshare{a})$ and $a \cdot G\from \Open(\gshare{a})$.
We use the command $ \Open $ to reconstruct from the secret shared values such that $a\from \Open(\sshare{a})$
For malicious security, we use message authentication code (MAC) scheme of SPDZ~\cite{cryptoeprint:2012:642,DBLP:conf/crypto/DamgardPSZ12}.

In SPDZ, a value $a$ is represented as $\sshare{a} = ((a_{1}, \dots, a_{N}), (\gamma(a)_{i}, \dots, \gamma(a)_{N}))$ where $ a_{i} $ is a share of $ a $ and $ \gamma(a)_{i} $ is the MAC share authenticating $ a $ under a global key $ \alpha\in\Zp $ such that $a=\sum_{i}a_{i}$ and $\alpha\cdot a = \gamma(a)=\sum_{i}\gamma(a)_{i}$.
Each party $i$ holds the pair $(a_{i},\gamma(a)_{i})$.
The execution of $ \Open $ in SPDZ involves the broadcast of the shares $ a_{i} $ by each party and computing $ \sum_{i} a_{i} $.
Then the MAC is checked to confirm that $ a $ is correct.
For this check, each party computes $\gamma_{i}(a)-\alpha_{i}a$, broadcasts the commitment and checks if $ \sum_{i}= \gamma_{i}(a)-\alpha_{i}a $.

Secure computing of ECDSA signatures does not only require the secret key  $ \sk $ to remain secret from all the parties but also the instance key $ k $.
The computation of $ k^{-1} $ should also be performed securely so that information about $ k $ is not revealed.
This is also the the most computationally expensive part of securely computing ECDSA signatures.

%\kris[inline]{High-level SPDZ, MASCOT, Shamir SS}

%\kris[inline]{
%%	SPDZ explanation\\
%	Add a one paragraph explanation of the threshold ecdsa solution\\
%	$ k^{-1} $ calculation.\\
%	No party should know $ k $ and $ \sk $.
%}

%% file: distributed_rpki.tex
\section{Distributed RPKI}
\label{sec:system}

\subsection{Threat Model}

\label{sec:system-threat-model}
\label{sec:threat-model}
%\kris[inline]{
%	What are the assumptions we make?\\
%	Which adversaries do we consider and which not?
%}

%In this section, we introduce the threat model in our system based on the threat models from MPC literature. Then, we introduce the system and communication model.

%\subsubsection{Threat Model}
%\label{sec:threat-model}

In our distributed RPKI system, we consider a stronger threat model than the existing RPKI system.
The existing threat model of RPKI includes external adversaries, but not the participating entities, such as RIRs, to be a possible attacker.
In this work, in addition to the threats considered in the existing system, we do not consider the RIRs to be entirely trustworthy.

Our threat model accounts for mistakes by the RIR as the hosted CA, the RIR under attack from an external adversary including legal coercion to modify, revoke or to inject RPKI data.
All these scenarios require access to the signing key for the attack to work.
We can capture these scenarios in our system by incorporating RIRs in the threat model.
Note that attacks on the publication point, such as deletion of RPKI data, are beyond the scope of this work.
%To be more accurate, we distribute trust so that individual RIRs do not need to be completely trusted.

%\kris[inline]{Outline the treat model in simple terms before introducing MPC terminology...link the two}
% What do we mean by trust here?
% It might be better to state it in simple words
% Trust in what and what not?

%We consider the guarantees that our distributed RPKI system provides.
Standard MPC terminology provides us with a tool kit to discuss threat models that not only includes external adversaries but also the participating parties.
Thus, we introduce standard MPC terminology to describe threat models in a distributed setting.
We consider adversary power, that is, whether an adversary is passive or active.
Then, we describe the guarantees that can be achieved when the threshold of honest parties varies.
Finally, we describe which guarantees our solution supports and how it translates to the threats against RPKI.

\paragraph{Honest-but-curious vs.~malicious security.}

MPC protocols can be classified in terms of the power of the adversary.
An adversary can be \emph{honest-but-curious} or \emph{malicious}.
An honest-but-curious adversary follows the protocol while a malicious adversary does not follow the protocol and might actively disrupt the protocol.
Security against honest-but-curious adversary is sufficient in many real life scenarios.
If the RIRs trust each other not to act maliciously and instead consider each other to be a necessary check on each other's operation, a protocol secure against honest-but-curious adversary is sufficient.
Such a protocol keeps the signing key away from any internal adversaries and curious employees at the RIRs.
On the one hand, a malicious adversary has full control over the RIRs and can disrupt the protocol.
For instance, it can send wrong values or delay the sending of values.
On the other hand, honest-but-curious adversaries are assumed to provide the correct inputs during the protocol so that no checks on the correctness of the inputs needs to be performed.
As might be obvious, a protocol against malicious adversaries provides stronger security guarantees.
%However, stronger security comes at the cost of efficiency.
%Not only is it harder but it is also less efficient to achieve security against malicious adversaries than against honest-but-curious adversaries.
%The difference in the efficiency depends on the specific protocol.

\paragraph{Honest vs.~dishonest majority.}

Let $n$ be the number of RIRs participating in the distributed RPKI system.
During the execution of a threshold signature protocol, a threshold $ t $ number of RIRs need to be available for the protocol to be successfully executed.
When a majority of the RIRs are honest, a threshold $ t \leq \lfloor (n-1)/2 \rfloor $ of parties are needed to sign, then it is called \emph{honest majority}.
When a minority of the RIRs are honest, that is, $ t < n $, the protocol is said to be secure for a \emph{dishonest majority}.
%Honest majority protocols are usually faster than dishonest majority protocols and are more beneficial when the availability of the distributed RPKI is of utmost importance.
%Honest majority protocols can be used even if some of the RIRs lose their key share as long as at least three-out-of-five RIRs are available and have their key share.
%Another way of looking at the security of honest majority protocols is as follows: as long as three-out-of-five RIRs do not collude, the key is not disclosed to anyone.
In the case of honest majority protocols, as long as three-out-of-five RIRs do not collude, the key is not disclosed to anyone.
As the RIRs often do not converge on the same policies, this may not be a strong assumption~\cite{Meul11}.
However, there are situations where a dishonest majority protocol might be needed as it provides stronger security such that the adversary needs to corrupt all the parties to be able to access the signing key.
A dishonest majority protocol might also be required when the signature should only be created when there is unanimity among the RIRs.
%However, dishonest majority protocols are vulnerable to unavailability when any one of the RIRs are unavailable or crash. 
%Hence, these two models provide an availability-security trade-off.

%\kris[inline]{Add reference to backup the claim that the RIRs do not collude is a reasonable assumption}

%\subsubsection{System and Communication Model}
%\label{sec:system-model}
%

\subsection{System setup}

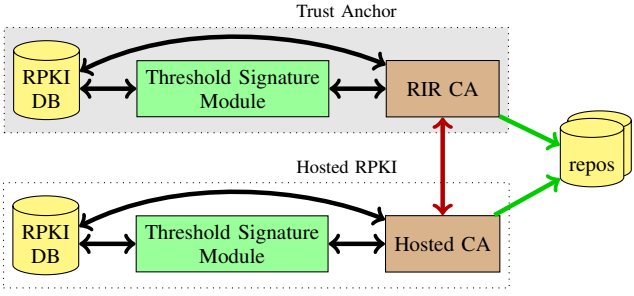
\begin{figure}[t!]
	\centering
	\input{Fig/Setup}
	\caption{System setup}
	\vspace{-10pt}
	\label{fig:setup}
\end{figure}

\begin{figure}[t!]
	\centering
	\input{Fig/Signing}
	\vspace{-10pt}
	\caption{Distributed RPKI architecture}
	\vspace{-10pt}
	\label{fig:distributed_setup}
\end{figure}
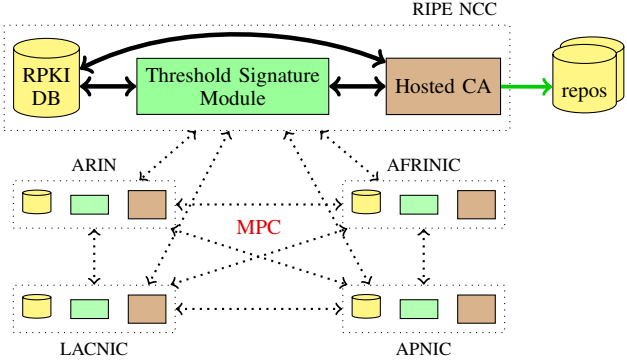

We present the system architecture in this section.
The setup at each RIR is shown in Figure~\ref{fig:setup} and the distributed RPKI architecture is shown in Figure~\ref{fig:distributed_setup}.
Each RIR has two components: Trust anchor and Hosted RPKI.
Each of them has a CA, a threshold signature module and access to a local RPKI database.
%\textbf{RIR CA}. The CA as part of the trust anchor, RIR CA, acts as a trust anchor in the RPKI.
%RIR CA is the one that issues the CA certificates to its members.
%It is also responsible to issue manifests and CRLs for the members.
%In addition, it also issues a self-signed certificate for itself and a certificate for the Hosted CA.
%RIR CA does not issue ROAs.
%Note that, although useful, it is not necessary to have two instances of the RIR CA---one online and another offline---to add redundancy to the system as is common in the existing RPKI systems as we distribute the RPKI process to increase availability.
%
%\textbf{Hosted CA}. The CA as part of the hosted RPKI, hosted CA, is the CA that is responsible to produce signed objects: ROAs, CRL and manifests for the members.
%It is available to all members of the RIR who choose to use Hosted RPKI.
%The public key and the share of the private key of the members is stored at the hosted CA.
%The members can access the hosted CA through a web interface, similar to those provided by the RIRs at present.
All the certificates and the signed objects issued by the two CAs are published in public access repositories, through rsync or RRDP~\cite{DBLP:journals/rfc/rfc8182}.

Our system incorporates all parts of the RPKI that requires generating signatures, which includes the creation of signed objects, ROAs, as well as signing of the resource certificates of children and the issuance of CRLs.
We distribute the task of creating signed objects among multiple RIRs.
As there are five RIRs, we use $ n = 5 $ in our system.

We focus on the key generation and the signing operation in a distributed RPKI system, such that no RIR has access to the signing key.
RIRs create their shares of the secret key and have access only to these share and not to the secret key.
For communication between the RIRs, we assume the existence of a synchronous communication network and that the protocols are run on a point-to-point network~\cite{DBLP:conf/focs/Canetti01,DBLP:conf/tcc/KatzMTZ13}.
We also assume that the RIRs use a secure and authenticated communication channel, e.g. using TLS.

%\textbf{Distributed RPKI architecture}. 
%Our system requires interaction between the RIRs for the creation of certificates and signed objects (Figure~\ref{fig:distributed_setup}).
%The RIRs are in different continents and the communication takes place over the public internet using secure and authenticated channels.

\subsection{DRPKI protocol phases}
\label{sec:proto_phases}

In our system, each RIR has a share of a private key for each member, uses this share to collaboratively issue signed objects and does not have access to the entire private key.
The protocols we use in this paper are in the security-with-abort model.
In this model, the protocol aborts if the participants are not available.
We note that threshold signature protocols that give guarantees on availability are possible.

The interactive protocols in our system are run between the 
All phases instigated by a Hosted CA and the interaction takes place between the Threshold Signature modules.
Due to the computation and communication overhead, we need an efficient threshold signing protocol.
This means, that when the signature is to be generated, there needs to be as little overhead as possible in comparison to traditional signatures.
This is possible with protocols that move most of the cryptography to the preprocessing phase and requiring minimum processing in the online phase, when the message to be signed is available.
Furthermore, we consider the efficiency in light of the threat models we discussed in Section~\ref{sec:threat-model}.
Based on these requirements, we adapt the threshold signature protocol of Dalskov et. al.~\cite{DBLP:conf/esorics/DalskovOKSS20} for our purpose (Figures~\ref{protocol:keygen}--\ref{protocol:sign}).

% \kris[inline]{Include a flow diagram key gen -> preprocessing -> offline data check (checking warrant or reason for CRL) -> Signing}

\subsubsection{Key generation}

\label{sec:proto_keygen}

In the key generation phase, new keys are generated such that each RIR generates a signing key share $ [\sk_j] $ for each member and runs the key generation protocol.
At the end of this phase, each RIR has the public key $ \pk_j $ and their share of the signing key $ \sshare{\sk_j} $.
The key generation protocol needs to be run every time keys are to be generated.
The keys do not need to be stored in a HSM.
The complete signing key is not exposed unless a threshold number (depending on the protocol being \emph{honest majority} or \emph{dishonest majority}) of RIRs have been compromised.
Figure~\ref{protocol:keygen} describes the protocol details.

\begin{Boxed}
	{Key generation $\kgen(\secparam)$}
	{protocol:keygen}
	{Key generation protocol}
	
	\begin{enumerate}
		\item Each RIR takes a security parameter $ \secparam $ as the input and generates a signing key share for the $ j^{th} $ member by randomly sampling $\sshare{\sk_{j}}\sample \Zp$.
		\item Each RIR locally converts $ \sshare{\sk_j} $ to $ \sshare{sk_j} \cdot G $.
		\item RIRs compute the public key \\ $\pk_j=\Open(\sshare{sk_j} \cdot G)) = \sk_j \cdot G $.
		\item Output the secret key shares and the public key $ (\sshare{\sk_{j}}, \pk_j) $.
	\end{enumerate}
\end{Boxed}

\subsubsection{Signing}

\label{sec:proto_signing}

The threshold signing protocol we use has two preprocessing phases and one online phase.
The first preprocessing phase is independent of the member for whom the signature is to be generated.
More specifically, this phase is independent of the signing key to be used.
This property allows us to amortize this phase.
This phase can be run between the RIRs before the member's request to generate a signature arrives.
Only an estimation of the number of signatures that would be required in a certain amount of time is required to run this phase.
At the end of this phase, the desired number of initial preprocessing tuples are generated and stored at each RIR.
Figure~\ref{protocol:preprop_indep} describes the protocol details.

\begin{Boxed}
	{Member independent preprocessing}
	{protocol:preprop_indep}
	{Member Independent preprocessing}
	
\begin{flushleft}{For each signature to be generated,}\end{flushleft}
\begin{enumerate}
		\item RIRs generate tuples of secret shared values of the form $(\sshare{a},\sshare{b},\sshare{c})$ such that $ a,b,c \in \Zp$ where $c=ab$.
		\item They open the share $ \sshare{c} $ by running \\ $c\from\Open(\sshare{c})$.
		\item Let $\sshare{k^{-1}}=\sshare{a}$.
		\item Each RIR locally generates $ \gshare{k} = (\sshare{b} \cdot G) \cdot c^{-1} $.
		%\item Then the RIRs compute  \\ $R\sample\Open(\sshare{b} \cdot G \cdot c^{-1}) = (bc^{-1})\cdot G = a^{-1} \cdot G = k \cdot G$.
		%\item Output the initial preprocessing tuple $(R,\sshare{k^{-1}})$.
		\item Output the initial preprocessing tuple $(\gshare{k},\sshare{k^{-1}})$.
		\vspace{-10pt}
\end{enumerate}
\end{Boxed}

The second preprocessing phase is dependent on the member for whom the signature is to be generated.
The threshold signature modules at the RIRs use the one unused initial preprocessing tuple.
It is security critical that the initial preprocessing tuples are not reused as it is equivalent to the reuse of the instance key $ k $.
%An attacker with two messages signed using the same signing key and instance key can recover the signing key.
At the end of this phase, the desired number of final preprocessing tuples are generated and stored at each RIR.
Figure~\ref{protocol:preprop_dep} describes the protocol details.

\begin{Boxed}
	{Member dependent preprocessing}
	{protocol:preprop_dep}
	{Member Dependent preprocessing}

\begin{enumerate}
		%\item RIRs take the generated signing key shares $\sshare{\sk_{j}}$ and the initial preprocessing tuple  $(R,\sshare{k^{-1}})$ as input.
		\item RIRs input the generated signing key shares $\sshare{\sk_{j}}$ and the initial preprocessing tuple  $(\gshare{k},\sshare{k^{-1}})$.
		\item They compute $\sshare{\sk_{j}'}=\sshare{\sk_{j}/k} $ by generating an additional tuple and Beaver's rerandomization technique~\cite{DBLP:conf/crypto/Beaver91a}.
		\item Output the final preprocessing tuple $(\gshare{k},\sshare{k^{-1}},\sshare{\sk_{j}'})$.
		%\item Output the final preprocessing tuple $(R,\sshare{k^{-1}},\sshare{\sk_{j}'})$.
\end{enumerate}
\end{Boxed}

In the final signing phase, the member gives consent to the changes that can be made through a standalone application.
This consent is sent to all the RIRs.
When a signature is to be generated, the message to be signed is sent by the RIR that initiates the signing protocol to the other RIRs.
%The message can take the form of ROAs, CRLs, CA certificates or end-entity certificates.
The message is checked, similar to the checks each RIR performs in the existing RPKI system, where they check the message locally for the message before they individually sign.
However, in our case, the check is performed by all the RIRs for all the messages that need to be signed.
Furthermore, the consent of the member is checked.
The RIRs check whether the consent has been given for the specific change, e.g., the transfer of IP space to another AS.
Note that a transfer of IP-space requires consent for a CRL for the existing EE certificate associated with the ROA and to create a new signed ROA.
These checks prevent RIRs to unilaterally take decisions to revoke certificates.
If a threshold number (depending on the protocol being \emph{honest majority} or \emph{dishonest majority}) of RIRs agree, then the RIRs locally compute their share of the signature before jointly computing the final signature.
Figure~\ref{protocol:sign} describes the protocol details.
With regards to the format of the messages, we do not make any change to the form and fields compared to the existing RPKI system.
The certificates take the form of X. 509 certificates~\cite{DBLP:journals/rfc/rfc6487} while the signed objects conform to RFC 6788~\cite{DBLP:journals/rfc/rfc6488}.
The RIRs check the contents of the message out-of-band.

\begin{Boxed}
	{Final signing phase}
	{protocol:sign}
	{Final signing phase}
	
	\begin{enumerate}
		\item The member uses a standalone application to give consent, e.g., to transfer IP-space to another AS. The consent is sent to all the RIRs.
		\item Input the message to be signed $ M $ and the final preprocessed tuple $(\gshare{k},\sshare{k^{-1}},\sshare{\sk_{j}'})$.
		%\item Input the message to be signed $ M $ and the final preprocessed tuple $(R,\sshare{k^{-1}},\sshare{\sk_{j}'})$.
		\item The RIR initiating the protocol sends the message $ M $ to the other RIRs.
		\item The RIRs check the contents of $ M $ and the consent by the member before proceeding. If the check fails, they abort $ \bot $. Else, they continue.
		\item Then the RIRs compute  \\ $R\sample\Open(\gshare{k}) = (bc^{-1})\cdot G = a^{-1} \cdot G = k \cdot G$.
		\item Let $(r_{x},r_{y})\leftarrow R$.
		\item Locally compute the share of the signature \\ $\sshare{s}=H(M)\cdot \sshare{k^{-1}} + r_{x}\cdot \sshare{\sk_{j}'}$.
		\item Finally compute the signature $s\from\Open(\sshare{s})$ and output $\sigma=(r_{x},s) $ or $ \sigma = \bot $.
		\vspace{-10pt}
		\end{enumerate}
\end{Boxed}

\subsubsection{Automation}

Our system is fully automated and does not need manual intervention in its regular operation.
Step 4 in Figure~\ref{protocol:sign} checks the message before it is signed.
This check, performed by all RIRs, verifies whether a consent has been received from the INR holder in Step 1.
If a threshold number of RIRs have not received the consent, then the check fails and the automated signing protocol aborts.

\subsubsection{Legitimate revocation without consent?}

So far, we have assumed that revocation of allocated IP resources requires the consent of the INR holder.
What about cases where there is a legitimate reason to revoke allocation?
Let us take a case where ARIN was fraudulently induced to issue IPv4 addresses~\cite{Arin19}.
After the fraud was detected and ARIN won a legal case, ARIN was able to take back the addresses.
Using our automated system with enforced consent, revocation of the IP address space in such a scenario will not be possible.
However, we are able to accommodate legitimate revocation with a minor change to the system.

Our automated system aborts the protocol if the check for consent fails at Step 4 in Figure~\ref{protocol:sign}.
Instead of aborting the protocol, we can flag it with the requirement for manual intervention if the protocol is to be completed.
Note that such a manual intervention will not require a large human effort as, in practice, most organizations obtain IP address space from their RIRs in good faith and there are only a few bad apples~\cite{RN19}.
We will require the RIRs to communicate off band before the protocol is completed.
This mechanism also allows for legitimate law enforcement requests to be processed by the RIRs, only when a threshold of other RIRs also agree.
Note that although technically possible, processing law enforcement mechanisms in this manner is akin to private regulation, which will require legal and policy changes for it to be realistic.

\subsection{Deployment scenarios}

%\kris[inline]{
%	System overview\\
%	How is this system wrt. to existing RPKI ecosystem?\\
%	What needs to be changed and what not?\\
%	What are the components of the system?\\
%	Is it easy to deploy? Why?\\
%	Are there assumptions that you make in this system?\\
%	Is it fully automated?\\
%	Security evaluation\\
%	If not, why? Which part requires human intervention?\\
%}

We propose two different deployment models.
We begin with a na\"{\i}ve deployment model and explain the reasons for its failure to solve the problem.
Then, we present two solutions with their associated trade-offs among the stake holders.
We emphasise that the trade-offs are not with respect to the security, but with respect to the responsibilities of the different stake holders.
%Finally, we discuss the network models that can be used to deploy our solution.

\textbf{Na\"{\i}ve solution:} In a na\"{\i}ve solution, our threshold signing module can be used for Hosted CAs.
This solution allows for the existence of the delegated CAs, which are beneficial to ISPs who sub-allocate resources.
This solution allows for IP-space owners to run their own CAs as well, that is, delegated CA system can continue to function in parallel with hosted CA system, as it does in the current system.
So the ISPs which have their own CA can delegate IP-space and sign the ROAs. 
However, only change is that the signing keys in the hosted CA setup are not in the possession of the individual RIRs.
The trust anchor from the existing RPKI exists and the RIR CA which is higher in the hierarchy can still revoke certificates unilaterally as it is not distributed.
And, the threat model remains weak and unchanged.

\textbf{Our solutions:} As the na\"{\i}ve deployment scenario does not solve the problem, we propose two deployment solutions.
Our first solution prevents unilateral takedown by the RIRs while our second solution also prevents LIRs from unilaterally taking down prefixes.
Both our solutions distribute the trust anchor.
Before discussing our solutions, we give an intuition behind our choice to distribute RPKI trust anchor.
The notion of a trust anchor requires all child nodes to unconditionally trust an entity.
In RPKI, there are five trust anchors, one at each RIR, which the relying parties use to verify RPKI signatures.
The concentration of power at trust anchors in the internet infrastructure extends beyond RPKI and is also observable in DNS(SEC) and Web PKI. Although many of the problems and vulnerabilities are similar~\cite{birge2018bamboozling,brandt2018domain,shulman2015towards,shulman2017one}, unlike DNS and Web PKI, there are already five trust anchors in RPKI that allows for a smooth transition to a distributed trust anchor.
Furthermore, the existing system of five trust anchors has had its issues.
As the policies of each RIR with regards to trust anchor is different, some relying parties do not use the trust anchor of ARIN and ROAs issued under ARIN's trust anchor locator fall to the status of `Not Found'~\cite{Tinka19}.
This means that even when RPKI is implemented, a significant portion of the networks do not validate routes originating from North America due to policy decisions and legal barriers~\cite{YW19}.
Thus, in practise large parts of the world are prevented from having better routing security.
These issues can be prevented if the trust anchor is not located at individual RIRs with their own policies and is instead distributed across them.

\textbf{Hierarchical deployment:}
In our first solution, we propose a two-layered tree deployment that maintains the hierarchical structure of RPKI.
In both layers, the RIRs use our threshold signature module.
The upper layer generates a distributed trust anchor to the five RIRs, while the lower layer uses the threshold signing module for the Hosted CAs.
In the upper layer, a distributed trust anchor is established using our key generation protocol in Figure~\ref{protocol:keygen}.
Each RIR generates their signing key share and participates in the key generation protocol to obtain the public key.
Once the public key is obtained, each RIR adds the public key to their TAL as the \texttt{subjectPublicKeyInfo}~\cite{DBLP:journals/rfc/rfc5280}.
Each RIR has a TA that has the same public key in the TAL.
As no RIR has the private key associated with this certificate, the RIR CAs do not need to be kept offline.
Thus, the RIRs do not need a subordinate CA to issue child certificates.
Furthermore, as each RIR has the same public key as part of the trust anchor and they have the same \texttt{subjectPublicKeyInfo} in their TAL, access to the TAL from one RIR is sufficient for relying parties to validate routes originating from any part of the world that has deployed RPKI.
Note that to generate the child certificates, RIRs run the signature generation protocol described in Section~\ref{sec:proto_signing}.

In the lower layer, our threshold signing module is used by the Hosted CAs to generate signed objects such as ROAs.
We are able to support delegated CAs as the distributed trust anchor at the RIR CAs is used to generate child certificates.
Furthermore, this solution allows for incremental deployment as the LIRs who have already deployed their own CAs can continue to use them to serve their child nodes while those who have not deployed their own CAs can start using hosted CA.
Note that the concerns regarding some LIRs being coerced by their country of registration remains.

\textbf{Flat deployment:}
In our second solution, instead of having the RIRs run two CAs, RIR CA and the Hosted CA, we combine the two so that the RIRs only need to run one CA.
Furthermore, we do not need a trust anchor as we replace the top-down architecture with a flat deployment architecture.
Not only do we eliminate the hierarchical structure of existing RPKI, we also distribute trust.
Moreover, this solution accounts for a stronger threat model where none of the CAs need to be completely trusted.
However, we do not support delegated CAs in this solution.
The CAs only generate end-entity certificates and signed objects; they do not generate any CA certificate that will allow child nodes to generate their own signed objects.
This also means that child nodes will need the RIRs to generate signed objects for their child nodes.
Nevertheless, we prevent any single entity to be all powerful and require the participation of a threshold number of RIRs for a signed object to be generated and ejected.

%% file: Fig/Setup.tex
%\documentclass[convert={convertexe={magick.exe},true}]{standalone}
%
%\usepackage{tikz}
%%usepackage[dvipsnames]{xcolor}
%\usetikzlibrary{backgrounds, calc, fit, positioning, shapes.geometric}

\pgfdeclarelayer{bg}
\pgfsetlayers{background,bg,main}

%\begin{document}
	\begin{tikzpicture}[
	scale=0.6,
	font=\footnotesize,
	box/.style={draw, rectangle, fill=brown!60, minimum height=0.75cm, minimum width=1.5cm},
	ebox/.style={draw, rectangle, fill=gray, minimum height=0.5cm, minimum width=1cm},
	control/.style={thick, <->},
	db/.style={draw, cylinder, minimum width=0.75cm, minimum height=0.5cm, fill=yellow!60!white, shape border rotate=90, aspect=.3, align=center},
	net/.style={draw, rectangle, dotted}
	]
	
	% Hosted RPKI
	
	\node[box] (ca) [align=center, shape aspect=1] {{Hosted CA}};
	\node[ebox] (tecdsa) [left=0.75cm of ca, align=center, shape aspect=1,fill=green!40] {Threshold Signature\\ Module};
	\node[db] (store) [left=0.75cm of tecdsa] {RPKI\\DB};
	
	\draw[control, line width=0.6mm,] (ca) -- (tecdsa);
	\draw[control, line width=0.6mm] (tecdsa) -- (store);
	\draw[control, line width=0.6mm] (store) .. controls ($ (store) !0.33! (ca) +(0,1.5) $) and ($ (store) !0.66! (ca) +(0,1.25) $) ..  (ca);

	\node[net] (op1) [minimum height=1.4cm, fit=(ca) (tecdsa) (store), label={[font=\scriptsize]60:Hosted RPKI}] {};

	% TrustAnchor
	\node[box] (ca2) [above=1.3cm of ca, shape aspect=1] {{RIR CA}};
	\node[ebox] (tecdsa2) [left=0.75cm of ca2, align=center, shape aspect=1,fill=green!40] {Threshold Signature\\ Module};
	\node[db] (store2) [left=0.75cm of tecdsa2] {RPKI\\DB};
	
	\draw[control, line width=0.6mm] (ca2) -- (tecdsa2);
	\draw[control, line width=0.6mm] (tecdsa2) -- (store2);
	\draw[control, line width=0.6mm] (store2) .. controls ($ (store2) !0.33! (ca2) +(0,1.5) $) and ($ (store2) !0.66! (ca2) +(0,1.25) $) ..  (ca2);
	
	\begin{pgfonlayer}{bg}
	\node[net] (op1) [minimum height=1.4cm, fill=lightgray!40, fit=(ca2) (tecdsa2) (store2), label={[font=\scriptsize]60:Trust Anchor}] {};
	\end{pgfonlayer}

	% external repo
	
	\node[db] (rsyncrepo0) [below right=0.05cm and 0.9cm of ca2] {\\repos};
	\node[db] (rsyncrepo) [below right=0.15cm and 0.8cm of ca2] {\\repos};

	\draw[control, line width=0.6mm, black!30!red] (ca) -- (ca2);
	\draw[control, line width=0.6mm, black!20!green, ->] (ca) -- (rsyncrepo);
	\draw[control, line width=0.6mm, black!20!green, ->] (ca2) -- (rsyncrepo);

	\end{tikzpicture}
%\end{document}

%% file: Fig/Signing.tex
%\documentclass[convert={convertexe={magick.exe},true}]{standalone}
%
%\usepackage{tikz}
%\usetikzlibrary{calc, fit, positioning, shapes.geometric}

%\begin{document}
	\begin{tikzpicture}[
	scale=0.6,
	font=\footnotesize,
	box/.style={draw, rectangle, fill=brown!60, minimum height=0.75cm, minimum width=1.5cm},
	ebox/.style={draw, rectangle, fill=gray, minimum height=0.5cm, minimum width=1cm},
	control/.style={thick, <->},
	db/.style={draw, cylinder, minimum width=0.75cm, minimum height=0.5cm, fill=yellow!60!white, shape border rotate=90, aspect=.3, align=center},
	net/.style={draw, rectangle, dotted}
	]
	
	% Hosted RPKI 1
	
	\node[box] (ca) [align=center, shape aspect=1] {{Hosted CA}};
	\node[ebox] (tecdsa) [left=0.75cm of ca, align=center, shape aspect=1,fill=green!40] {Threshold Signature\\ Module};
	
	\node[db] (store) [left=0.75cm of tecdsa] {RPKI\\DB};
	
	\draw[control, line width=0.6mm,] (ca) -- (tecdsa);
	\draw[control, line width=0.6mm] (tecdsa) -- (store);
	%\draw[cond] (c1) .. controls ($ (c1) !0.33! (v2) +(0.75,0) $) and ($ (c1) !0.66! (v2) +(0.5,0) $) .. (v2);
	%\draw[control, line width=0.6mm, blue] (store) -- (ca);
	\draw[control, line width=0.6mm] (store) .. controls ($ (store) !0.33! (ca) +(0,1.5) $) and ($ (store) !0.66! (ca) +(0,1.25) $) ..  (ca);
	
	\node[net] (op1) [minimum height=1.4cm, fit=(ca) (tecdsa) (store), label={[font=\scriptsize]20:RIPE NCC}] {};
	
	% Hosted RPKI 2
	
	\node[box] (ca2) [below left = 1cm and -0.4cm of tecdsa, minimum height=0.375cm, minimum width=0.5cm, shape aspect=1] {};
	\node[ebox] (tecdsa2) [left=0.25cm of ca2, align=center, shape aspect=1,fill=green!30,minimum height=0.25cm, minimum width=0.5cm] {};
	
	\node[db] (store2) [left=0.25cm of tecdsa2, minimum width=0.375cm, minimum height=0.25cm] {};
	
	\node[net] (op2) [minimum height=0.5cm, fit=(ca2) (tecdsa2) (store2), label={[font=\scriptsize]90:ARIN}] {};

	% Hosted RPKI 3
	
	\node[box] (ca3) [below= 1cm of ca2, minimum height=0.375cm, minimum width=0.5cm, shape aspect=1] {};
	\node[ebox] (tecdsa3) [left=0.25cm of ca3, align=center, shape aspect=1,fill=green!30,minimum height=0.25cm, minimum width=0.5cm] {};
	
	\node[db] (store3) [left=0.25cm of tecdsa3, minimum width=0.375cm, minimum height=0.25cm] {};
	
	\node[net] (op3) [minimum height=0.5cm, fit=(ca3) (tecdsa3) (store3), label={[font=\scriptsize]270:LACNIC}] {};

	% Hosted RPKI 4
	
	\node[box] (ca4) [below right = 1cm and 1.7cm of tecdsa, minimum height=0.375cm, minimum width=0.5cm, shape aspect=1] {};
	\node[ebox] (tecdsa4) [left=0.25cm of ca4, align=center, shape aspect=1,fill=green!30,minimum height=0.25cm, minimum width=0.5cm] {};
	
	\node[db] (store4) [left=0.25cm of tecdsa4, minimum width=0.375cm, minimum height=0.25cm] {};
	
	\node[net] (op4) [minimum height=0.5cm, fit=(ca4) (tecdsa4) (store4), label={[font=\scriptsize]90:AFRINIC}] {};

	% Hosted RPKI 5
	
	\node[box] (ca5) [below= 1cm of ca4, minimum height=0.375cm, minimum width=0.5cm, shape aspect=1] {};
	\node[ebox] (tecdsa5) [left=0.25cm of ca5, align=center, shape aspect=1,fill=green!30,minimum height=0.25cm, minimum width=0.5cm] {};
	
	\node[db] (store5) [left=0.25cm of tecdsa5, minimum width=0.375cm, minimum height=0.25cm] {};
	
	\node[net] (op5) [minimum height=0.5cm, fit=(ca5) (tecdsa5) (store5), label={[font=\scriptsize]270:APNIC}] {};

	% external repo
	
	\node[db] (rsyncrepo0) [above right =-0.6cm and 0.9cm of ca] {\\ repos};
	\node[db] (rsyncrepo) [above right =-0.7cm and 0.8cm of ca] {\\ repos};
	\node (MPC) at ($ (op2) !.5! (op4) + (0,-0.5)$) {\textcolor{black!10!red}{MPC}};

	\draw[control, dotted] (op1) -- ($ (op2.north) + (1,0) $);
	%($(store1.south) + (0.5,0) $)
	\draw[control, dotted] (op1) -- ($ (op3.north) + (1.2,0) $);
	\draw[control, dotted] (op1) -- ($ (op4.north) + (-1,0) $);
	\draw[control, dotted] (op1) -- ($ (op5.north) + (-1.2,0) $);

	\draw[control, dotted] (op2) -- (op3);
	\draw[control, dotted] (op2) -- (op4);
	\draw[control, dotted] (op2) -- (op5);
	
	\draw[control, dotted] (op3) -- (op4);
	\draw[control, dotted] (op3) -- (op5);
	
	\draw[control, dotted] (op4) -- (op5);
	
	\draw[control, line width=0.5mm, black!20!green, ->] (ca) -- (rsyncrepo);

%	
%	
%	%	
%	\draw[control, line width=1mm, black!30!red] (ca) -- (ca2);
%	
%	%\draw[control, ->] (ca) -- node[below right, pos=0.65, align=center] {\scriptsize rir-ta.cer, rir-ta.mft,\\ \scriptsize  rir.crl, hostedca.cer} (rsyncrepo);
%	\draw[control, line width=1mm, black!20!green, ->] (ca) -- (rsyncrepo);
%	\draw[control, line width=1mm, black!20!green, ->] (ca2) -- (rsyncrepo);

	\end{tikzpicture}
%\end{document}

%% file: eval.tex
\section{Implementation and evaluation}

\label{sec:eval}

We have implemented our system in C++ and have used MP-SPDZ~\cite{mp-spdz} for the threshold ECDSA MPC protocols.
MP-SPDZ includes threshold ECDSA protocol implementations for all the security models that we are concerned with: honest-but-curious and malicious as well as honest and dishonest majority protocols.
In particular, we use four protocols---Shamir, Mal. Shamir, Semi OT and MASCOT---that are shown in Table~\ref{tab:MPC_proto}.
The former two are based on Shamir secret sharing while the latter two are based on additive secret sharing.
We use all four protocols to implement the system described in Section~\ref{sec:proto_phases}.

\begin{table}
	\centering
	\footnotesize
\begin{tabular}{c|c|c}
	%\hline 
	\backslashbox{Majority}{Adversary power} & Honest-but-curious & Malicious  \\ 
	\hline 
	Honest  & Shamir & Mal. Shamir \\ 
	\hline 
	Dishonest  & Semi. OT & MASCOT  \\ 
	\hline 
\end{tabular} 
  \caption{Four MPC protocols}
  \vspace{-10pt}
\label{tab:MPC_proto}
\end{table}

\subsection{Deployment Setups}
\label{deployment}

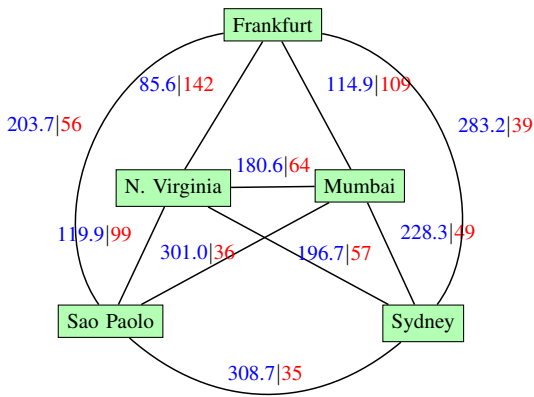
\begin{figure}[!ht]
	\centering
	\input{Fig/latency}
	\caption{\textcolor{blue}{Latency}$ \mid $\textcolor{red}{Bandwidth} between regions, where latency is in milliseconds and bandwidth is in Mbits/s}
	\label{fig:latency}
	\vspace{-10pt}
\end{figure}

For performance evaluation, two deployments were set up.
For each node, we used an Amazon AWS c5.2xlarge instance with a 64-bit Intel Xeon CPU with 3 GHz and 16 GB RAM.
We run all the evaluations on a single thread.
To make our evaluations as realistic as possible, we chose to run the experiments based on the location of the RIRs. % (See Table~\ref{tab:location}).
The five RIRs are in different continents of the world.
So, in the first setting, we run experiments on five Amazon AWS instances that are placed around the world such that they are representative of the location of the RIRs.
Specifically, we use the instances at Frankfurt, N.Virginia, Sydney, Sao Paolo and Mumbai while the RIRs are based in Amsterdam, Virginia, Brisbane, Sao Paolo, Mauritius, respectively.
The latency and the bandwidth between the instances is shown in Figure~\ref{fig:latency}.
Furthermore, we also consider the setting where the RIRs could, in the future, have virtual servers located close to other RIRs.
For this purpose, we also run our experiments on the LAN in Frankfurt.

%\begin{table}
%	\centering
%	%\small
%\begin{tabular}{c|c|c}
%	%\hline 
%	\textbf{RIR} & \textbf{RIR location} & \textbf{AWS location} \\ 
%	\hline 
%	AFRINIC & Mauritius & Mumbai \\ 
%	\hline 
%	APNIC & Brisbane & Sydney \\ 
%	\hline 
%	ARIN & Virginia & N. Virginia \\ 
%	\hline 
%	LACNIC & Sao Paolo & Sao Paolo \\ 
%	\hline 
%	RIPE NCC & Amsterdam & Frankfurt \\ 
%	\hline 
%\end{tabular} 
%  \caption{Location of RIRs and AWS instances}
%\label{tab:location}
%\end{table}

\subsection{Experimental evaluations}

\textbf{Key generation.}
We benchmark the 5-party key generation protocol in both the settings.
The total key generation time is composed of the timings for generating secret key and public key.
Secret key generation involves generating a field element $ \sshare{sk} $] while public key generation involves a local conversion of the field element into an elliptic curve point of order before being opened.
The timings shown in Table~\ref{tab:keygen} are the mean and standard deviation over 10 executions of the protocols where the value taken for each execution is the time noted when the last party completes the protocol.
While the honest majority protocols (Shamir and Mal. Shamir) only require one round of communication for secret key generation, dishonest majority protocols (Semi OT and MASCOT) are costlier, especially in WAN setting.

\begin{table*}[h]
	\centering
	\begin{tabular}{lrrr|rrr}
		\toprule
		& \multicolumn{3}{c}{LAN} & \multicolumn{3}{c}{WAN}	\\
		\cmidrule(r){2-4} \cmidrule(r){5-7}
		& Secret & Public & $ \kgen $ 
		& Secret & Public & $ \kgen $ 
		\\
		\midrule
		MASCOT       & $ 6.99 \pm 0.04 $ & $ 2.48 \pm 0.02 $ & $ 9.47 \pm 0.03 $ & $ 4490 \pm 1.74 $ & $ 1147 \pm 0.27 $ & $ 5637 \pm 1.25 $\\
		Semi OT      & $ 0.88 \pm 0.04 $ & $ 0.91 \pm 0.02 $ & $ 1.79 \pm 0.03 $ & $ 851 \pm 2.53 $ & $ 486 \pm 0.93 $ & $ 1337 \pm 1.91 $  \\
		Mal. Shamir  & $ 0.24 \pm 0.00 $ & $ 1.59 \pm 0.03 $ & $ 1.83 \pm 0.02 $ & $ 198 \pm  0.23 $ & $ 487 \pm 0.61 $ & $ 685 \pm 0.46 $ \\
		Shamir       & $ 0.25 \pm 0.04 $ & $ 1.13 \pm 0.08 $ & $ 1.38 \pm 0.06 $ &  $ 284 \pm 1.38 $ & $ 382 \pm 3.48 $ & $ 666 \pm 2.64 $ \\
		
		\bottomrule
	\end{tabular}
	\caption{Breakdown of key generation timings in milliseconds for $\sshare{\sk}$ sharing, and $\pk$.}
	
	\label{tab:keygen}
\end{table*}

\begin{table*}[h]
	\centering
	\small
	\begin{tabular}{lrrr|rrr}
		\toprule
		& \multicolumn{3}{c}{LAN} & \multicolumn{3}{c}{WAN}	\\
		\cmidrule(r){2-4} \cmidrule(r){5-7}
		& Preprocessing & Online & $ \sig $ 
		& Preprocessing & Online & $ \sig $ 
		\\
		\midrule
		MASCOT       & $ 4.78 \pm 0.01 $ & $ 1.89 \pm 0.02 $ & $ 6.67 \pm 0.02 $ & $ 50.56 \pm 1.86 $ & $ 1055 \pm 37.23 $ & $ 1106 \pm 26.36 $ \\
		Semi OT      & $ 0.96 \pm 0.01 $ & $ 1.51 \pm 0.01 $ & $ 2.47 \pm 0.01 $ & $ 9.00 \pm 0.90 $ & $ 487 \pm	0.40 $ & $ 496 \pm 0.70 $ \\
		Mal. Shamir  & $ 1.43 \pm 0.00 $ & $ 1.40 \pm 0.02 $ & $ 2.83 \pm 0.01 $ & $ 10.94 \pm 0.68 $ & $ 283 \pm 0.06 $  & $ 294 \pm 0.48 $  \\
		Shamir       & $ 0.98 \pm 0.00 $ & $ 1.30 \pm 0.02 $ & $ 2.28 \pm 0.01 $ & $ 3.77 \pm 0.00 $ & $ 282 \pm 0.18 $ & $ 286 \pm 0.13 $ \\
		
		\bottomrule
	\end{tabular}
	\caption{Breakdown of signing timings in milliseconds for preprocessing and online phases per signature. Preprocessing times are based on amortized generation of 1000 tuples.}
	\label{tab:signing_time}
\end{table*}

\textbf{Signing.}
We benchmark the preprocessing time (member dependent and independent) to generates tuples and the online signing time per signature in Table~\ref{tab:signing_time}.
For preprocessing, we present the time taken to generate one tuple when 1000 tuples are generated in an amortized manner.
As the preprocessing does not depend on the message to be signed, thousands of preprocessed tuples can be generated and stored.
They can be used when a new message is to be signed.
Note that the online phase does not involve any elliptic curve operation and, hence, is computationally cheap.

Although dishonest majority protocols are generally costlier than honest majority protocols, Semi OT has the highest preprocessing throughput in LAN setting (Table~\ref{tab:signing_throughput}).
Semi OT protocol uses additive sharing which is cheaper than elliptic curve operations, which is the predominant cost during preprocessing.
In the WAN setting communication becomes more predominant than local operations.
We also observe that the cost of malicious security in the case of honest majority protocol is very small.
This is especially true in the WAN setting as the extra checks for Mal. Shamir are local operations while communication becomes the predominant cost.

In Table~\ref{tab:comm}, we show the communication per party for the four protocols.
We note that the communication is asymmetric for Mal. Shamir and Shamir.
Hence, we present the mean of the communication over all the parties.
We notice that the preprocessing communication per tuple as well as online signing is significantly higher for dishonest majority protocols than honest majority protocols.
In comparison, the communication overhead per party is marginal for malicious security over honest-but-curious protocols.

\begin{table}[h]
	\centering
	\footnotesize
	\begin{tabular}{lrr|rr}
		\toprule
		& \multicolumn{2}{c}{LAN} & \multicolumn{2}{c}{WAN}	\\
		\cmidrule(r){2-3} \cmidrule(r){4-5}
		& Preprocessing & Online 
		& Preprocessing & Online 
		\\
		\midrule
		MASCOT       & $ 209 $ & $ 529 $ & $ 20 $ & $ 0.95 $ \\
		Semi OT      & $ 1042 $ & $ 662 $ & $ 111 $ & $ 2.05 $ \\
		Mal. Shamir  & $ 699 $ & $ 714 $ & $ 91 $ & $ 3.53 $ \\
		Shamir       & $ 1020 $ & $ 769 $ & $ 265 $ & $ 3.54 $ \\
		
		\bottomrule
	\end{tabular}
	\caption{Breakdown of throughput for preprocessing (tuples/sec) and online phases (signatures/sec).}
	\label{tab:signing_throughput}
\end{table}

\begin{table}[t]
	\centering
	\footnotesize
	\begin{tabular}{lrrr}
		\toprule
		& $ \kgen $ & Preprocessing (per tuple) & Online Signing 
		\\
		\midrule
		MASCOT      & $ 0.482 $ & $ 624 $	& $ 0.400 $ \\
		Semi OT     & $ 0.113 $ & $ 99.0 $ & $ 0.128 $ \\
		Mal. Shamir & $ 0.271 $ & $ 1.345 $ & $ 0.0768 $ \\
		Shamir      & $ 0.206 $ & $ 0.437 $ & $ 0.0512 $\\
		
		\bottomrule
	\end{tabular}
	\caption{Communication per party (KByte)}
	\label{tab:comm}
\end{table}

%% file: Fig/latency.tex
\newcommand{\ping}[1]{\color{blue}{#1}\normalcolor}
\newcommand{\bw}[1]{\color{red}{#1}\normalcolor}

\begin{tikzpicture}[
scale=0.6,
font=\footnotesize,
loc/.style={draw, rectangle, fill=green!30},
control/.style={thick, line width=0.2mm, -},
control1/.style={arc, line width=0.2mm, -}
]

\node[loc] (fra) [align=center, shape aspect=1] {Frankfurt};
\node[loc] (syd) [below right = 3.5cm and 0.8cm of fra] {Sydney};
\node[loc] (sao) [below left = 3.5cm and 0.8cm of fra] {Sao Paolo};
\node[loc] (mum) [below right = 1.7cm and -0.1cm of fra] {Mumbai};
\node[loc] (vir) [below left = 1.7cm and -0.1cm of fra] {N. Virginia};

\draw[control] (fra) -- node[above right, pos=0.5] {\ping{114.9}$ \mid $\bw{109}} (mum);
\draw[control] (fra) -- node[above left, pos=0.5] {\ping{85.6}$ \mid $\bw{142}} (vir);

\draw[control] (syd) -- node[above right, pos=0.5] {\ping{228.3}$ \mid $\bw{49}} (mum);
\draw[control] (syd) -- node[above, pos=0.3] {\ping{196.7}$ \mid $\bw{57}} (vir);

\draw[control] (sao) -- node[above left, pos=0.5] {\ping{119.9}$ \mid $\bw{99}} (vir);
\draw[control] (sao) -- node[above, pos=0.3] {\ping{301.0}$ \mid $\bw{36}} (mum);
\draw[control] (mum) -- node[above, pos=0.5] {\ping{180.6}$ \mid $\bw{64}} (vir);

\draw[control] (fra) .. controls ($ (fra) !0.33! (sao) +(-2.7,1.5) $) and ($ (fra) !0.66! (sao) +(-2.8,0) $) .. node[above left, pos=0.5] {\ping{203.7}$ \mid $\bw{56}} (sao);
\draw[control] (fra) .. controls ($ (fra) !0.33! (syd) +(2.7,1.5) $) and ($ (fra) !0.66! (syd) +(2.8,0) $) .. node[above right, pos=0.5] {\ping{283.2}$ \mid $\bw{39}} (syd);
\draw[control] (sao) .. controls ($ (sao) !0.33! (syd) +(0,-2) $) and ($ (sao) !0.66! (syd) +(0,-2) $) .. node[above, pos=0.5] {\ping{308.7}$ \mid $\bw{35}} (syd);
%\draw[control] (sao) to[out=325,in=215] node[above, pos=0.5] {x} (syd);

\end{tikzpicture}

%% file: analysis.tex
\vspace{-5pt}
\section{Analysis}

\label{sec:analysis}

%\kris[inline]{
%Use history of RPKI data, check the frequency of changes (addition/removal) of ROAs and plot the information.\\
%Source of data?\\
%What is the purpose of this analysis?\\
%What is the inference from this analysis?\\
%Is it reasonable to extrapolate the numbers under the condition that all the prefixes are signed?\\
%}

For the deployment of our distributed RPKI system, it needs to be efficient enough.
In the previous section we discussed the efficiency in terms of the runtime of our protocols.
In this section, we discuss whether they are efficient enough in terms of the number of signatures required by the RIRs.
As our system involves all the five RIRs, we take into account the cumulative of the requirements of all of them.

\paragraph{RPKI data}

We accessed the publicly available historical RPKI data maintained by RIPE NCC that includes the daily archive of the repositories of all the five RIRs from 2011 onwards~\footnote{\url{https://ftp.ripe.net/rpki/}}.
%We use the historical data from 1 March 2015 till 19 February 2020.
We use the historical data from 11 March 2015 till 10 August 2020.

%\kris[inline]{Update the exact date before submission.}
%\kris[inline]{Is the data from the past few months enough?}
%\kris[inline]{Include a table for the total number of ROAs per RIR on a specific date; e.g., 19 Feb}
%\kris[inline]{- For CRL and resource certificates, CA key (only one needed)\\
%	- For ROAs of children, many keys are to be generated.}
%\kris[inline]{Public data set}

\paragraph{ROA analysis}

\begin{figure*}[!ht]
	\begin{minipage}{.5\textwidth}
		\centering
		\includegraphics[width=0.8\linewidth]{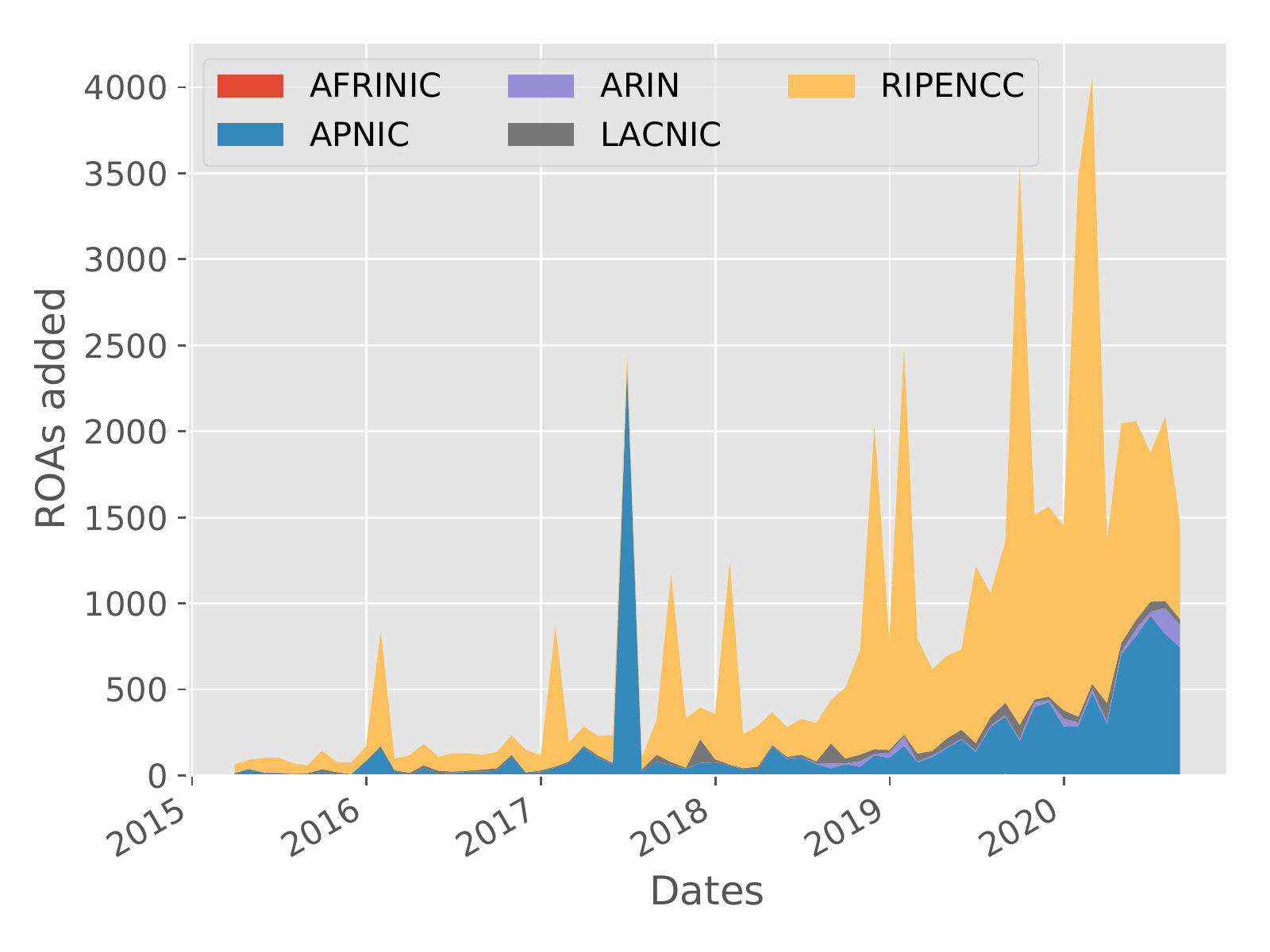}
		%\caption{ROAs added}
		%\label{fig:plotroaaddedarin}
	\end{minipage}%
	\begin{minipage}{.5\textwidth}
		\centering
		\includegraphics[width=0.8\linewidth]{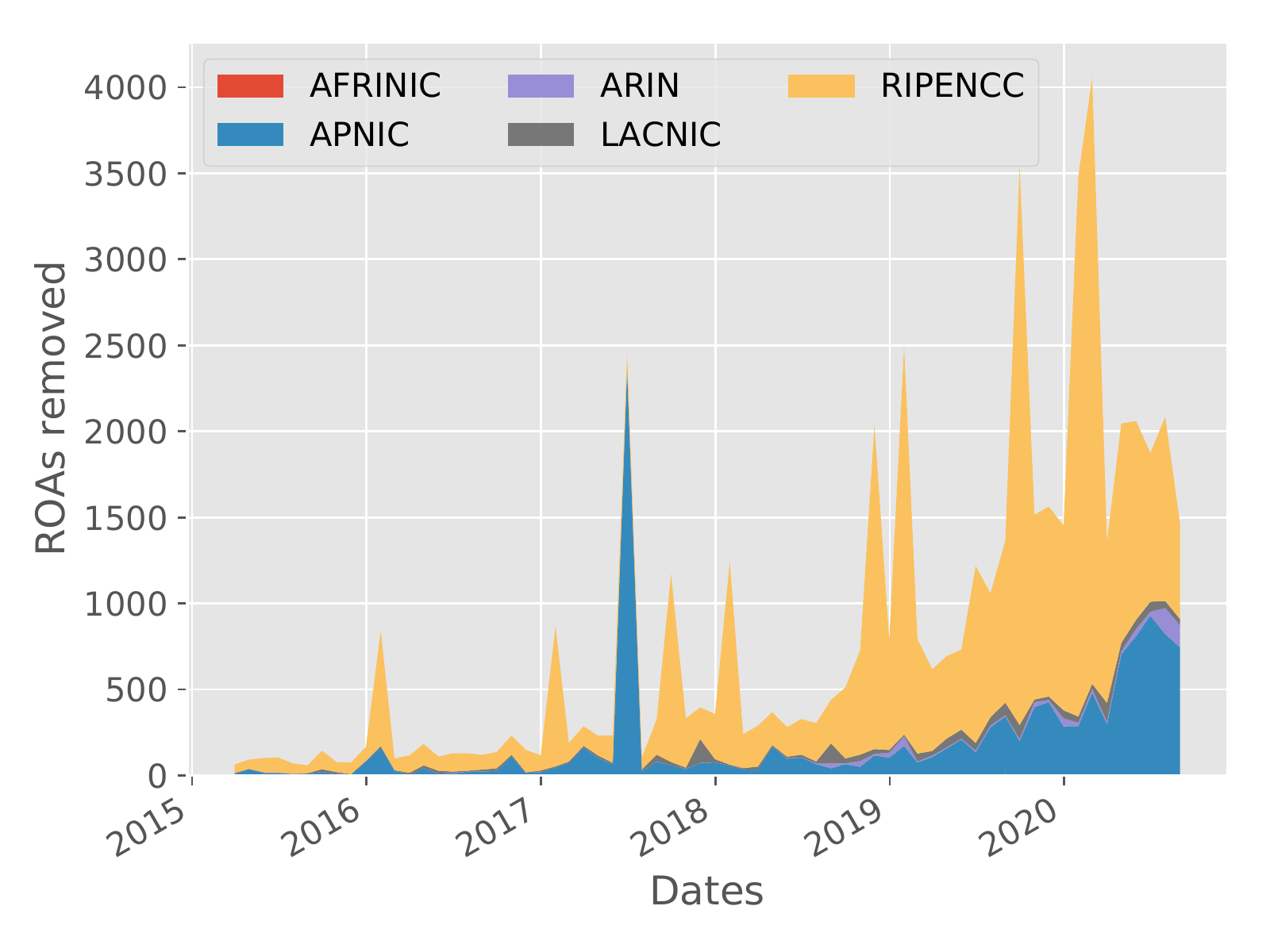}
		%\caption{ROAs removed}
		%\label{fig:plotroaremovedarin}
	\end{minipage}%
	\caption{Number of ROAs added and removed on average per day from March 2015 to August 2020}
	\vspace{-10pt}
	\label{fig:mean_add_remove_ROAs_2015_20}
\end{figure*}

\begin{figure*}[!ht]
	\begin{minipage}{.5\textwidth}
		\centering
		\includegraphics[width=0.8\linewidth]{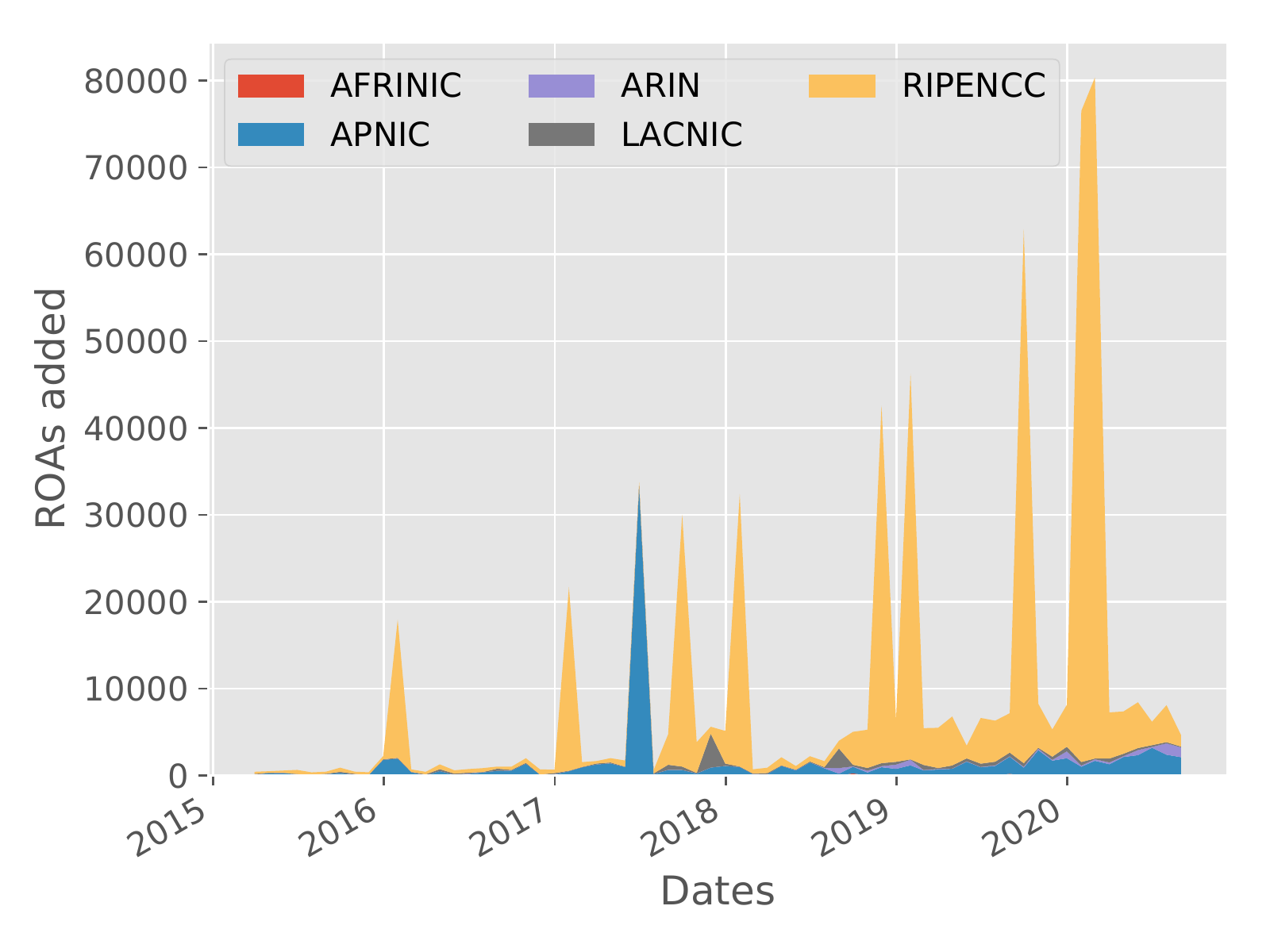}
		%\caption{ROAs added}
		%\label{fig:plotroaaddedarin}
	\end{minipage}%
	\begin{minipage}{.5\textwidth}
		\centering
		\includegraphics[width=0.8\linewidth]{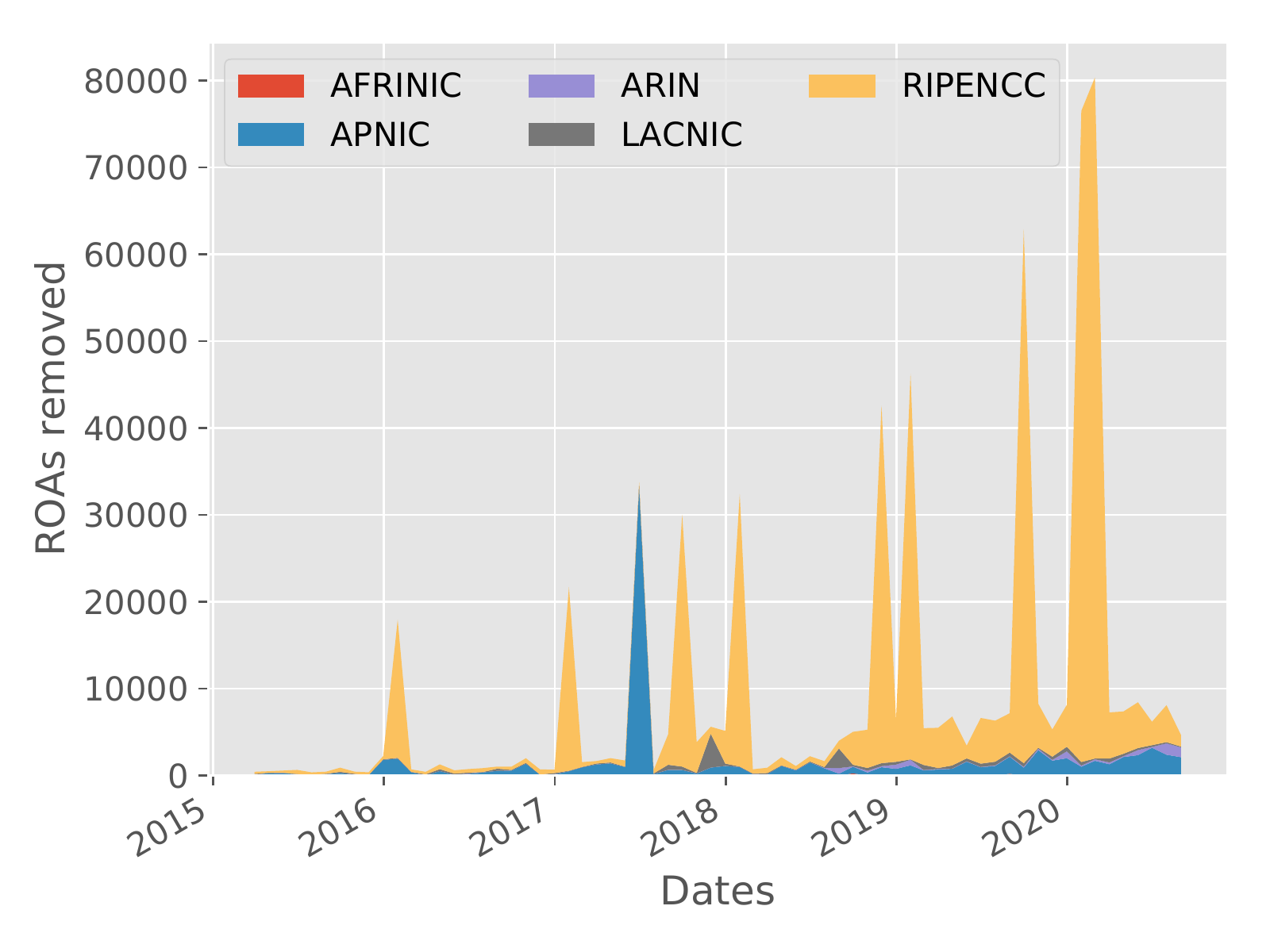}
		%\caption{ROAs removed}
		%\label{fig:plotroaremovedarin}
	\end{minipage}%
	\caption{Maximum per month of ROAs added and removed from March 2015 to August 2020}
	\label{fig:max_add_remove_ROAs_2015_20}
	\vspace{-10pt}
\end{figure*}

We use the RPKI data to analyse the number of ROAs that have been added and removed per day in a certain time period.
We estimate the number of signatures required based on this information.
Figure~\ref{fig:mean_add_remove_ROAs_2015_20} shows the change on average day (mean taken over a month) in ROAs for the five RIRs. % while Figure~\ref{fig:add_remove_ROAs_2019_20} shows the daily change in the first two months of 2020.
On average, we need about $ 8000 $ signatures per day.
However, there are days when the load is greater.
This occurs on days when many ROAs are re-issued.
Figure~\ref{fig:max_add_remove_ROAs_2015_20} shows the maximum number of changes per month.
Note the scale on y-axis: There is a twenty times difference from Figure~\ref{fig:mean_add_remove_ROAs_2015_20}.

%However, there are exceptions, such as 1 January 2020, when there was a factor 10 increase change in the ROAs.
We observe from Table~\ref{tab:signing_throughput} that for our slowest protocol MASCOT, we are able to produce $ 0.95 $ signatures/sec or $ 82080 $ signatures/day in the WAN setting.
For our fastest protocol, we are able to produce $ 3.54 $ signatures/sec or $ 305856 $ signatures/day in the WAN setting.
Even our slowest protocol can produce 10x more signatures than is required on an average day.
All our other protocols are fast enough even on days with peaks in Figure~\ref{fig:max_add_remove_ROAs_2015_20}.
In the LAN setting, all our protocols are fast and have the capacity to produce three orders of magnitude more signatures.
The efficiency of our system makes it possible to scale it as the adoption of RPKI increases.

%\kris[inline]{Check the new numbers and update}

%- For removal, one CRL per CA, i.e., one signature needed.

%\kris[inline]{A graph showing the increase in ROAs per RIR could be useful as well.}

%\kris[inline]{%Think about alternative representation of the graph to make it more intuitive\\
%	%Plots for a short durations\\
%	%Use stacked plots instead of line\\
%	Explain the graphs\\
%	The changes are not consistent. There are some tall peaks. Is MPC efficient enough able to cover such events as well? Explain in the next section.\\
%	Explanation for the peaks will be useful to the reader. Are these anomalies or misconfiguration?\\
%	Add a graph (CDF) including all RIRs in one as well}

%% file: relatedworks.tex
\section{Related Work}

\label{sec:relatedworks}

%\kris[inline]{We build on the proposal in ANRW'20 and construct a full system and perform extensive evaluations.\\
%	In this paper, we go further by constructing and describing a full system along with extensive performance evaluations.\\
%	%limitations?
%}

%\subsection{Threshold Signatures for Internet Infrastructure.}

{\bf Threshold Signatures for Internet Infrastructure.} Threshold signatures for DNSSEC have been considered in the past.
Cachin and Sanar~\cite{cachin2004secure} proposed a distributed DNS to avoid single point of failure, %which provides fault tolerance and security in the presence of corrupted servers. 
while Cifuentes et. al~\cite{DBLP:conf/lanc/CifuentesHMBRB16} use threshold signatures to emulate a HSM at an authoritative name server.
Dalskov et. al~\cite{DBLP:conf/esorics/DalskovOKSS20} use threshold signatures to secure DNSSEC signing keys when the zone and key management is outsourced to DNS operators. Integration of new algorithms into DNSSEC can be done with ciphersuite negotiation mechanisms and similar ideas can be applied for RPKI \cite{herzberg2014less,herzberg2014negotiating,herzberg2015cipher}. Finally, Shrishak and Shulman~\cite{DBLP:conf/anrw/ShrishakS20} initiated the research direction of using threshold signatures for RPKI. We extend that work further by developing and setting up a complete system in the Internet along with extensive performance evaluations.

%\subsection{Limiting the power of RIRs}

{\bf Limiting the power of RIRs.} There have been two approaches in the prior works to limit the power of RIRs in RPKI.
One approach has been to add transparency logs and \texttt{.dead} objects to RPKI to note the consent of the Internet Number Resource (INR) owner for revocation~\cite{DBLP:conf/sigcomm/HeilmanCRG14}.
Heilman et. al~\cite{DBLP:conf/sigcomm/HeilmanCRG14} detect when a ROA has downgraded from valid to invalid or valid to unknown state and check whether a \texttt{.dead} object is present. This approach requires relying parties to perform ROVs, if it is to be effective.
ROV deployment monitor~\footnote{\url{https://rov.rpki.net}}, a monitoring platform~\cite{rbcks-trmma-18} shows that only a few relying parties, %about 100 ASes, 
124 ASes, perform ROV, while there are %68289 ASes as of 1 April 2020
67599 ASes as of 10 August 2020~\footnote{\url{https://www.caida.org/data/as-relationships/}}.
%Furthermore, ROAs have error and hence, are not trusted by relying parties.
%(2)~It performs detection while we attempt to prevent malicious activities.
%Attacks can be detected after the fact due to the transparency and the effort of the relying parties.
%Accountability in the form of malicious activity detected post-mortem is not sufficient as ASes may have already lost out on their business.
Furthermore, this approach fails in the hosted RPKI setting as the \texttt{.dead} objects that are used to signify consent from the child  can be signed by the parent node allowing the parent to impersonate the child.

The second approach replaces the existing RPKI system with blockchain~\cite{DBLP:conf/hotnets/HariL16}.
In this approach, the role of RIRs is limited to providing new resources and they cannot revoke already allocated resources.
In addition to the large-scale changes that this approach requires, blockchain has other deployment issues such as consensus algorithm and the lack of incentive for the nodes to run the blockchain.
While \cite{DBLP:conf/ithings/PaillisseFGLPLK18} suggested to use proof-of-stake as the consensus algorithm, it has the possibility to create another form of power imbalance where the nodes with a larger stake such as large ISPs will become powerful players.

%However, both these approaches do not account for the cases where there might be a legitimate reason to revoke the allocated IP-address blocks, e.g., due to fraud by INR owner~\cite{Arin19}.
%It should be possible to revoke allocations under legitimate circumstances, but not otherwise.

%As blockchain-based proposals suffer from scalability issues, RouteChain~\cite{DBLP:conf/icbc2/SaadAAAYM19} employs a hierarchy of ASes which are assigned in subgroups to validate BGP announcements on the blockchain.
%However, in practice, ASes may have conflicting policies that prevent dynamic grouping and, thus, an incentive mechanism is also required.

% Blockchain requires a change in how RPKI is governed by RIRs which makes it a hard sell. Why should RIRs consider a solution which reduces their influence? In our solution, RIRs still have an influence but individual RIRs cannot make unilateral decisions.

%\kris[inline]{- Compare with active and passive threats.\\
%- detection and prevention approach\\
%- ease of deployment\\
%- Limitations of related work}

%% file: conclusion.tex
\section{Conclusion}

\label{sec:conclusion}

RPKI offers security benefits to BP and yet, it is not widely deployed.
One reason is that it opens up the possibility for unilateral IP-prefix takedown.
In this work, we make a small change to RPKI and propose a distributed RPKI that relies on prevention rather than detection of takedowns.
As our solution requires communication between the RIRs, we hope that it will re-instigate discussions between the RIRs on the need for further collaboration.
We propose two deployment models, the second of which eliminates the hierarchical structure of the existing RPKI and flattens the power relations.
Both our deployment models distribute the trust anchor and prevent the scenario where validation fails due to the unavailability of a trust anchor.
We perform extensive evaluation to assess the efficiency of our solution based on four threshold signature protocols and show that our solution scales when the deployment of RPKI increases.

%We note that our solution is purely technical, and legal and policy barriers need to be addressed to make the work truly practical.

%% file: acknowledgement.tex
\section*{Acknowledgment}

This work has been co-funded by:
the German Federal Ministry of Education and Research and the Hessen State Ministry for Higher Education, Research and Arts within their joint support of the National Research Center for Applied Cybersecurity ATHENE; %the DFG as part of project D.3 within the RTG 2050 “Privacy and Trust for Mobile Users" and as part of project S3 within the CRC 1119 CROSSING; 
the Deutsche Forschungsgemeinschaft (DFG, German Research Foundation): GRK~2050/251805230 and SFB~1119/236615297.